\documentclass[twocolumn,aps,showpacs,prl]{revtex4}

\usepackage{graphicx,latexsym,endnotes,color}
\usepackage{psfrag,amsmath,epstopdf}
\usepackage[english]{babel}
\graphicspath{{figures/}}

\begin{document}

\author{M. Caraglio}
\email{caraglio@pd.infn.it}
\affiliation{Dipartimento di Fisica e Astronomia, Sezione INFN, CNISM, 
Universit\`a di Padova,
 Via Marzolo 8, I-35131 Padova, Italy}
\author{A. Imparato}
\email{imparato@phys.au.dk}
\affiliation{Department of Physics and Astronomy, University of Aarhus,
  Ny Munkegade, Building 1520, DK--8000 Aarhus C, Denmark}
  
\title{Energy Transfer in molecular devices}

\date{\today}

\begin{abstract}
Protein machines often exhibit long range interplay between different sites in order to achieve their biological tasks.
We investigate and characterize the non--linear energy localization and the basic mechanisms of energy transfer in protein devices. By studying  two different model protein machines, with different biological functions, we show that genuinely non--linear phenomena are responsible for energy transport  
 between the different machine sites involved in the biological functions.  
The energy transfer turns out to be  extremely efficient from an energetic point of view:
by changing the energy initially provided to the model device, we identify a well defined range of energies where 
the time for the energy transport to occur is minimal and the amount of transferred energy is maximum. Furthermore, by introducing an implicit solvent, we show that the energy is localized on the internal residues of the protein structure, thus minimizing the dissipation.
\end{abstract}

\maketitle

\begin{figure}[h]
\center
\includegraphics[width=4.25cm]{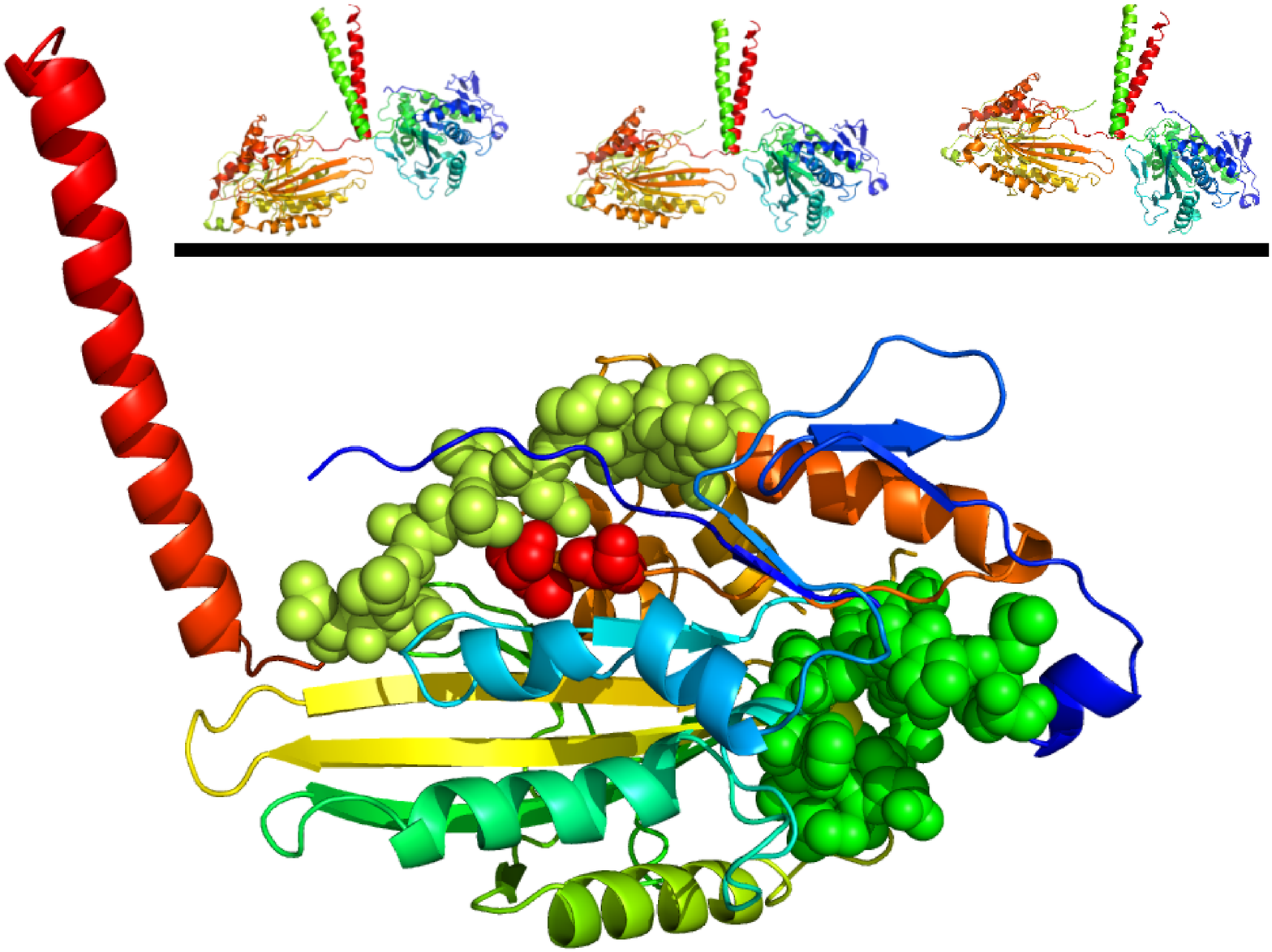}
\includegraphics[width=4.25cm]{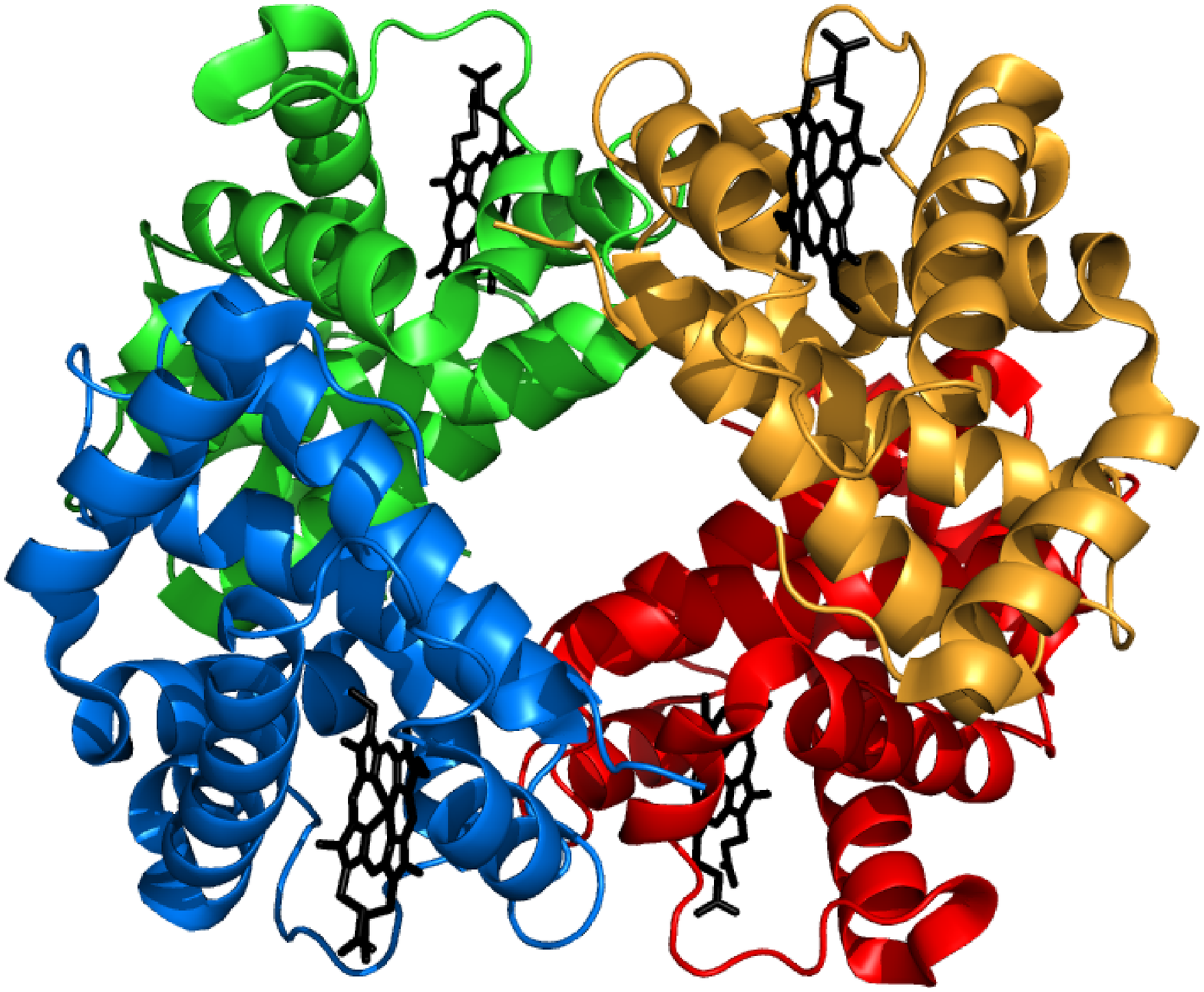}
\caption{Left: cartoon representing the two headed molecular motor kinesin which hydrolyzes ATP molecules to walk on a microtubule. The two heads are connected via a short, flexible neck linker to the stalk, a long, central alpha-helical coiled-coil domain (red or green $\alpha$-helix). Bottom: Particular of the single head structure, the ATP binding site (dark green spheres), the neck linker (light green spheres) and two residues in the neck linker docking site (CYS 296 and THR 298, red spheres) are highlighted. Right: Structure of the hemoglobin molecule (PDB code 2HHB) with the four heme groups inside each of the four subdomains (black). Each heme group binds one oxygen molecule. In the text, we indicated the four subdomains with the letters  A,B,C, and D, starting from the bottom--right and proceeding in the counterclockwise direction.
 }
\label{fig:kin}
\end{figure}

Molecular devices in cells, such as enzymes and molecular motors,  are microscopic thermodynamic systems that convert one form of energy, typically chemical energy, into useful work to
accomplish several tasks such as chemical reaction catalysis, intracellular transport, protein and nucleic acids synthesis, ion pumping across cell membranes, and cell locomotion~\cite{Alberts2002,Nelson2008,Howard2001}. 
Most of the molecular devices in cells are made of proteins, protein assemblies, or are complexes of proteins with other molecules. Indeed
proteins are versatile biopolymers which fold into a compact globular structure  
whose shape is strictly connected to their biological functions.
Differently from their
macroscopic counterparts, these devices operate at constant temperature, and given their microscopic size, they are subject to large thermal fluctuations. 
However, similarly to macroscopic machines, the energy needed for the functioning is produced in sections of the structure not immediately in contact with the sites where this energy is used to perform useful work. In macroscopic devices this problem is solved  with a system of shafts, pulleys and transmitting belts.  In molecular motors, such as kinesin (Fig.~\ref{fig:kin} left), the energy produced in a catalytic site has to be transmitted across the molecular structure up to the point where it is employed to perform mechanical work, and thus produce movement.
This process must necessarily involve some  collective conformational changes in the protein device.
It is widely known that protein dynamics is highly anharmonic \cite{Levy82,Hayward95}, and thus one must consider non-linear effects when investigating proteins functioning as devices. Indeed, it has been shown that, as a consequence of spatial disorder and nonlinear interaction between the residues, phenomena  of energy storage and transmission across the protein structure can emerge~\cite{PiazzaPRL2007,Piazza2008,PiazzaEPL2009,PiazzaPB2009,Luccioli2011}. In particular, the Discrete Breathers (DBs) have attracted a lot of attention. They are spatially localized, time-periodic modes, which are able to harvest most of the energy in a protein structure, and concentrate it on few sites~\cite{PiazzaPRL2007, Piazza2008,PiazzaEPL2009,PiazzaPB2009,Luccioli2011}, in particular they are able to confine energies larger that the typical thermal energy on a few protein residues, for long times, and are able to pump energy across a protein, several nanometers away from where the energy is released~\cite{PiazzaPB2009,Luccioli2011}.
All the recent studies in the field have concentrated on DBs formation and energy transmission in proteins. More specifically it has been highlighted that DBs concentrate the background energy on the  stiffest region of the protein backbone~\cite{Piazza2008}.
Furthermore, a systematic analysis carried out on a large set of enzymes \cite{PiazzaPRL2007,Piazza2008} has shown that DBs mostly form in  the catalytic regions.
In the present letter we show that these non-linear effects have a central role in protein machine functioning. We study two molecular machines, the kinesin motor and the hemoglobin,  the oxygen carrier in red blood cells of all vertebrates (Fig.~\ref{fig:kin} right),  and show that both of them exhibit non--linear energy localization and transfer between the relevant sites involved in the device biological function. We argue that such an energy transfer is an efficient mechanism, both without and with solvent.
Our findings strongly support the hypothesis that localized vibrational modes of non-linear origin play a central role in biological processes~\cite{Peyrard95,Peyrard08}. 
We show that these  effects are particularly important in allosteric control of protein devices~\cite{Nelson2008}, where the binding of one ligand to one specific site affects the properties of other sites: both the devices we consider are indeed typical examples  of allosteric proteins.

We use here the Nonlinear Network Model (NNM) introduced in ref.~\cite{PiazzaPRL2007} to study energy transfer phenomena in proteins.
In the NNM a $N$--residue protein is described by a sequence of $N$ coarse-grained particles with the same mass ($m = 110$ gr/mol), and position at equilibrium determined by the coordinates of the corresponding C{$_{\alpha}$} atoms as found in the protein data bank (PDB)~\cite{pdb}.
Given the distance $d_{ij}$ between residues $i$ and $j$ the potential energy reads
\begin{equation} \label{Eq_PotentialEnergy}
U = \sum_{i=1}^N \sum_{j=1}^N c_{ij} \left[ \dfrac{k_2}{4} (d_{ij}-d^0_{ij})^2 + \dfrac{k_4}{8} (d_{ij}-d^0_{ij})^4 \right] \; ,
\end{equation}
where $k_2$, $k_4$ are two constant parameters and $c_{ij}$ is a contact matrix element which takes the value $c_{ij}=1$ if $d^0_{ij} \le d_c$ and $c_{ij}=0$ otherwise, where $d^0_{ij}$ is the distance between the residues in the native configuration, and $d_c$ is a  specific cut--off distance.

In accordance with previous studies~\cite{PiazzaPRL2007,Piazza2008,PiazzaEPL2009,PiazzaPB2009} we take $k_2 = 5$ Kcal/mol/\AA$^2$, $k_4 = 5 $ Kcal/mol/\AA$^4$ and $d_c=10$ \AA . For $k_4= 0$, the NNM reduces to the  Elastic Network Model (ENM) which, despite its simplicity, captures the essential dynamics of  amino-acid fluctuations at room temperature~\cite{Tirion1996,Bahar1997} and have been used to investigate the large amplitude motions of proteins upon ligand binding~\cite{Delarue2002}. 
Nevertheless, we emphasize that in the present paper we use the NNM to investigate the energy localization and transfer phenomena, which occur on time scales much shorter than larger conformational changes.

We first perform Normal Mode (NM) Analysis for our protein devices, by linearizing the potential energy (\ref{Eq_PotentialEnergy}) around the equilibrium position and  finding the eigenvalues and the eigenvectors of the Hessian matrix~\cite{PiazzaPRL2009}.

To study the energy transfer phenomena, we perform microcanonical molecular dynamics simulations recording both the single site and NM energies.
We label the NMs with the index $k$, and define $\omega_k$ and $\pmb{\xi}^k$ the corresponding eigenvalues and eigenvectors of the Hessian matrix.
NM energy $\epsilon_k(t)$ at time $t$ is evaluated as $\epsilon_k(t)=(\dot{Q}^2_k + \omega_k^2 Q_k^2)/2$ where  $Q_k = \sum_{j,\alpha} m (r_{j,\alpha}-r_{j,\alpha}^0) \xi^k_{j,\alpha} \;$, with $r_{j,\alpha}$ the residue coordinates and $r_{j,\alpha}^0$ their value in the equilibrium configuration ($j=1,2,\ldots,N$; $\alpha=x,y,z$).
Following \cite{Piazza2008, PiazzaPB2009} we initialize our simulations  in two different ways.
For the relevant sites, we find the maximal strain direction (MSD), as given by the Sequential Maximum Strain (SMS) algorithm~\cite{Piazza2008}. Simply put, this amounts to finding the direction around a residue where the local energy landscape is steepest. We then initialize by giving a {\it kinetic kick} to the given residue, i.e., by assigning an initial kinetic energy $E_0$ along the MSD. 
Another possibility is to directly excite one of the NMs with a sufficient amount of initial kinetic energy \cite{PiazzaPB2009}, i.e., given the $k$-th NM, we assign to the $j$-th residue an initial velocity $v_{j,\alpha} \propto \xi^k_{j,\alpha}$ ($\alpha=x,y,z$), with the proportionality factor such that the system total energy takes the desired value $E_0$.

We start our analysis with the kinesin motor. It is made up of a head domain with $\sim 340$ amino acid residues and dimension $7\times 4.5\times4.5$ nm~\cite{Vale1996} whose radius of gyration is about $19.6$ \AA, a soft neck linker, and a central stalk (the neck)~\cite{Kozielski1997}, see Fig.~\ref{fig:kin}.

Residues 16-19, 87-94, 233-239 and 198-202 are thought to be involved in ATP binding~\cite{Kozielski1997}.
Experimental evidence~\cite{Nelson2008,Rice1999,Miyazono09} indicates that the neck linker is mobile and becomes more ordered when the catalytic core binds an ATP molecule.
The ATP binding induces a conformational change that docks the neck linker region alongside its head, advancing the stalk together with the partner head.
After the release of the hydrolysis byproducts (ADP+P), the neck linker returns to its mobile conformation. Here we show that the non-linear effects cause the energy to be pumped from the kinesin catalytic core to the neck linker docking site.
Thus, there is an interplay between the energy produced at the catalytic site and the allosteric change of the neck-linker, which leads to the overall movement of the motor. 

\begin{figure}[h]
\center
\includegraphics[width=8cm]{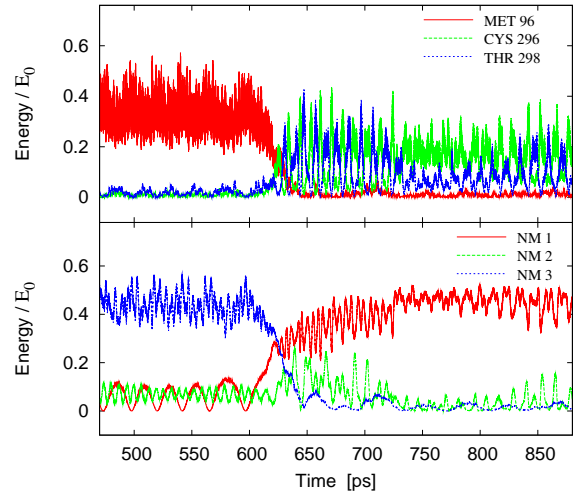}
\caption{(color online) Energy transfer in Kinesin after a kick of energy $E_0 = 30$ Kcal/mol at site MET 96. Upper panel: energy vs. time of MET 96, CYS 296 and THR 298. Lower panel: energy vs. time of NM1, NM2 and NM3.}
\label{Fig_transfer}
\end{figure}

In Fig.~\ref{Fig_transfer} we show a long-range energy transfer event occurring in Kinesin from Rattus Norvegicus (PDB code 3KIN) which is initially given a kinetic energy kick of magnitude $E_0=30$ Kcal/mol at site MET 96, very close to the ATP pocket. In the upper panel the energy of sites MET 96, CYS 296 and THR 298 is reported as a function of time, while the lower panel shows the energy of the three NMs with the highest frequencies, NM 1, 2 and 3 respectively. The energies associated with all the other residues (or NMs) are smaller than a few percent of the total energy during the whole simulation. Thus, depending on the time, one of the three residues (or one of three first NMs) is also the most energetic residue (NM). 
According to a standard metric to express the degree of localization of a NM, which depends on the pattern of the corresponding eigenvector $\pmb{\xi}^k$, NM3 is mainly localized on MET 96 \footnote{
A NM $k$ is said to be localized if only few residues vibrate with an appreciable amplitude
when the specific NM is excited.
Thus NM $k$, with eigenvector components $\xi^k_{i,\alpha}$ ($i=1,\ldots,N \, ; \, \alpha = x,y,z$), is localized on residue $i$ if the amplitude of the vibrations of residue $i$ due to NM $k$ is much greater than the amplitudes of all the other residues: namely if $ \sum_{\alpha} (\xi^k_{i,\alpha})^2  \gg \sum_{\alpha} (\xi^k_{j,\alpha})^2 \; , \; \forall j \ne i$. Similarly, NM $k$ is said to be localized on both residue $i_1$ and $i_2$ if
$\sum_{\alpha} (\xi^k_{i_1,\alpha})^2 \simeq \sum_{\alpha} (\xi^k_{i_2,\alpha})^2 \gg \sum_{\alpha} (\xi^k_{j,\alpha})^2 \; , \; \forall j \ne i_1,i_2$.
}.
To a lesser extent, NM3 is also localized on the whole ATP pocket region, so our initial conditions are intended to reproduce the typical excitation occurring immediately after the ATP hydrolysis.
At very short time, the residue  MET 96 loses  about $60$\% of the total energy mainly towards those residues involved in NM3 fluctuations.
Thus, given the initial condition that we have chosen the energy is initially mostly localized on MET 96 and consequently on NM3.
Because of the quartic term in Eq.~(\ref{Eq_PotentialEnergy}) NM3 loses energy towards other modes. Initially this drop is very slow but at time $t \approx 600$ ps the energy of NM3 quickly decreases in favour of the first two NMs, completing the energy transfer. In particular, after such a transfer, NM1 becomes the most energetic mode. NM1 is highly localized on the two residues CYS 296 and THR 298 which  together carry approximately $40$\% of the total energy after the transfer.
Note that these two sites are about $20$ \AA ~away from the one kicked at the beginning and they are located in the region where the neck linker docks, thus this simple model might account for the fist step in of the mechanism leading to the forward movement of the head as described above~\cite{Rice1999}, by pumping energy from the ATP binding site to the neck-linker region.
Kicking other residues in the ATP pocket, either singularly or in group, leads to similar results: if energy transfer takes place, it always flows towards CYS 296 and THR 298 (data not shown).
However the details of the dynamics such as the amount of transferred energy and time at which the transfer occurs depend strongly on the initial conditions.




We evaluated the typical frequency of the localized mode, before and after the energy transfer has taken place, through Principal Component Analysis (PCA)~\cite{Hayward2008}.
It consists in evaluating the mass-weighted velocity covariance matrix  as given by $m \langle v_{i,\alpha} v_{j,\beta} \rangle$, where $m$ is the residue mass, $v_{i,\alpha}$ is the $\alpha$ component of $i$-th residue's velocity and $ \langle \cdot \rangle$ denotes a time average over a timespan along a simulated trajectory.
The eigenvectors of such a matrix are called Principal Modes (PMs) and, as for NMs, are usually sorted with their eigenvalues in decreasing order.
Each eigenvector corresponds to a given correlated motion while its eigenvalue provides the average kinetic energy of that motion.
 
With reference to the simulated trajectory shown in Fig.~(\ref{Fig_transfer}), the PCA performed from $0.4$ ns to $0.6$ ns shows that NM3 is the mode the closer to PM1, the scalar product of the corresponding eigenvectors being equal to $0.82$, while the Fourier spectrum of the system trajectory projected onto PM1 shows that it has a single dominant frequency which is approximately $2$\% higher than NM1 frequency.
On the other hand, PM1 evaluated during the post-transfer dynamics, from $0.75$ ns to $1$ ns, has a substantial superimposition with the NM1,  the eigenvector scalar product being equal to $0.99$, and a single dominant frequency approximately $3$\% higher than NM1 frequency.
This aspects, together with the energy transfer, signals that we are in presence of genuinely non--linear effects, and the localized mode we observe are DBs which are nonlinear continuation of NM3, before the transfer, and of NM1, after the transfer~\cite{PiazzaPB2009}.


\begin{figure}[h]
\center
\includegraphics[width=8cm]{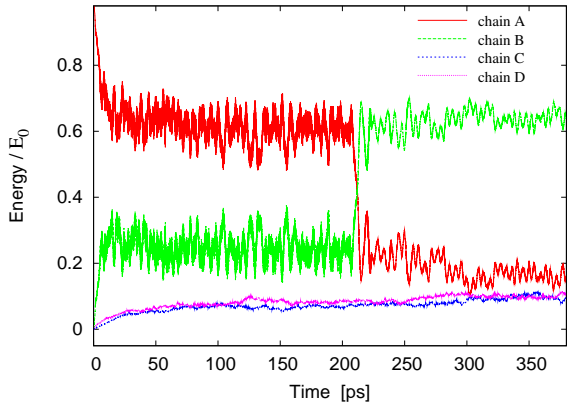}
\caption{(color online) Energy transfer in Hemoglobin. Chains energy vs. time after an initial kick of energy $E_0 = 50$ Kcal/mol at site SER 102 in chain A.}
\label{Fig_transfer_hemo}
\end{figure}

Another relevant example of allosteric protein is the hemoglobin, see Fig.~\ref{fig:kin}, right.
Indeed, after binding one oxygen molecule, hemoglobin is predisposed to bind more because of the long range interaction between its four binding sites~\cite{Eaton99}.
We find that, similarly to the kinesin motor, Hemoglobin (PDB code 2HHB) exhibits a non linear energy transfer and localization between its binding sites.
We kicked one of the binding sites of the heme group in chain A of this protein, namely SER 102, according to the SMS algorithm with $E_0=50$ Kcal/mol. As shown by Fig.~\ref{Fig_transfer_hemo}, the energy is quickly shared between chain A ($\approx 60$\%), chain B ($\approx 20$\%), chains C and D ($\approx 10$\% each). This energy partition is unchanged up to time $t\approx 210$ ps, with the most energetic residue being VAL 107 in chain A. Then chain A suddenly transfers about $60$\% of its energy to chain B and the most energetic residues becomes ALA 27 in chain B: this non--linear energy transfer might be the basic mechanism leading to the increased binding affinity in the second binding site, after an oxygen molecule has bound the first binding site. 
The reverse energy transfer from chain B to chain A can be observed if one provides initial energy to chain B. Similarly, if the energy is given to the binding sites in chain C(D), it is then transferred to chain D(C).
In the controversy between concerted and sequential models~\cite{Ackers2006} regarding the functional mechanism of hemoglobin, our results are consistent with sequential models according to which 
 after the first oxygen binding, the changes in the binding affinity of the other subdomains occur as a cascade of events.

\begin{figure}[h]
\center
\includegraphics[width=8cm]{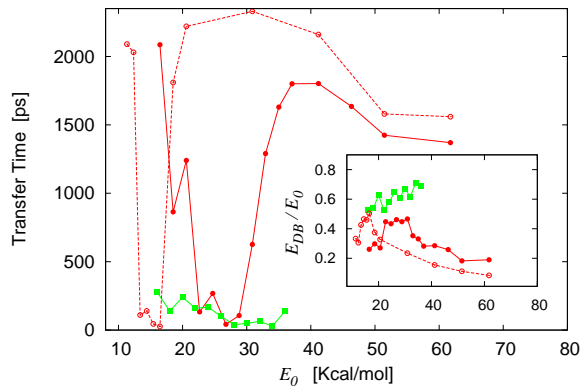}
\caption{(color online) Transfer time in Kinesin as a function of initial energy $E_0$. Red line--circles: initial kinetic energy kick at site MET 96 according to the SMS algorithm, with $k_2 = 5$ Kcal/mol/\AA$^2$, $k_4 = 5 $ Kcal/mol/\AA$^4$ (full line) and $k_2 = 5$ Kcal/mol/\AA$^2$, $k_4 = 8.5 $  Kcal/mol/\AA$^4$ (dashed line). Green line--squares: initial kinetic energy kick at the NM3, $k_2 = 5$ Kcal/mol/\AA$^2$, $k_4 = 5 $ Kcal/mol/\AA$^4$. Inset: energy of the localized mode after the energy transfer has occurred, full lines $k_2 = 5$ Kcal/mol/\AA$^2$, $k_4 = 5 $, dashed line $k_2 = 5$ Kcal/mol/\AA$^2$, $k_4 = 8.5 $ Kcal/mol/\AA$^4$.}
\label{Fig_transfer_efficiency}
\end{figure}

In order to investigate how 
the characteristics of the initial energy kick influence the transfer efficiency, we repeated the above analysis for kinesin with different values of $E_0$, either giving energy to certain residues according to the SMS algorithm or directly feeding specific NMs.
Extra care must be paid since in the present model, given the contact matrix and the distances between various atoms, the energy landscape displays frustration and even the configuration with zero potential energy may not be uniquely defined. Indeed,  the system energy as given by eq.~(\ref{Eq_PotentialEnergy}) depends on the residue distance and not on their actual position, so different configurations may exhibit the same energy.
NM analysis may be affected if the supplied energy is high enough to move the system from the correct equilibrium configuration to another local minimum. Furthermore, if the energy is too high, spurious dynamical evolution may arise because of round-off errors in the force computation during the simulation. To overcome such a problem, we upper bounded $E_0$ by checking that, in the absence of the quartic term in Eq.~(\ref{Eq_PotentialEnergy}), every NM keeps its initial energy along the whole trajectory.
Fig.~\ref{Fig_transfer_efficiency} shows  both the transfer time (defined as the time at which transfer process is completed) and  the fraction of transferred energy into localized mode, as  functions of the initial energy $E_0$ given to the kinesin. When directly feeding the NM3 (green line--squares) energy transfer is relatively fast for all the values of $E_0$ below the bound discussed above with the localized DB collecting about $60$\% of the total energy. 
While the transfer efficiency slightly increases with the initial energy for the range of $E_0$ considered here, ($E_0\le 36$ Kcal/mol, see discussion above).
This behaviour is not monotonous, in a simulation with $E_0=40$ Kcal/mol (data not shown) we find that the transfer has not yet occurred after $2$ ns. Remarkably, when the simulation is initialized by giving a single kick to the site MET 96 along the MSD  (red full line--circles), the energy transfer is very fast and energetically efficient for $E_0$ ranging from $22$ to $28$ Kcal/mol while the process is slower and inefficient outside this optimal interval.
While we have chosen the value for $k_2$ and $k_4$ following previous works in this field \cite{PiazzaPRL2007,Piazza2008,PiazzaEPL2009,PiazzaPB2009}, we find that  by decreasing the ratio $k_2/k_4$ (see fig.~\ref{Fig_transfer_efficiency}, dashed line) one observes a shift of the curve (full line circles)   such that the optimal transfer (minimal transfer time and maximal transfer energy) is obtained for values of $E_0$ closer  to the known value for the energy released by hydrolyzing one ATP molecule ~\cite{Howard2001} ($\sim 10$ Kcal/mol).


\begin{figure}[htbp]
\center
\includegraphics[width=8cm]{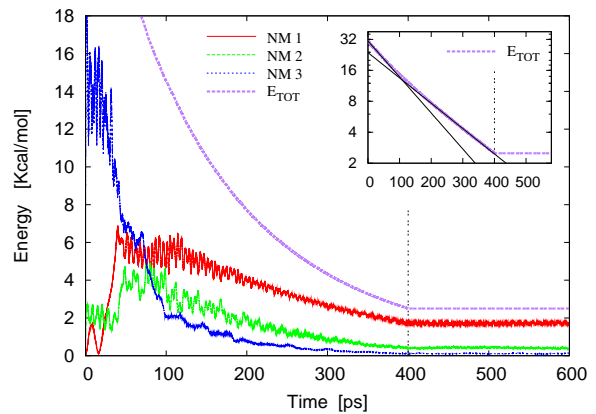}
\caption{(color online) Formation of DB in Kinesin after an initial kinetic kick of energy $E_0 = 30$ Kcal/mol at site MET 96. Energy of NM1, NM2, NM3 and total energy vs. time. The surface friction acts up to $400$ ps ($\gamma = 0.2$ ps$^{-1}$) and then it is turned off. Inset: energy exponential decay $E_{\mbox{\scriptsize{TOT}}} \propto e^{-x/\tau}$; black solid lines are fit from $0$ to $50$ ps and from $200$ to $400$ ps.}
\label{Fig_Friction}
\end{figure}

Finally since proteins interact and exchange energy with the surrounding environment through their surface, it is reasonable to investigate how our results change when considering an implicit solvent. For this purpose, following previous investigations~\cite{PiazzaPRL2007,Luccioli2011,AubryPRL1996,Zambrini2013},  we introduce a viscous friction for residues which can be reached by a test solvent particle (with radius $1.4$ \AA) rolling on the protein surface. The surface dissipation is known to stabilize the most localized modes~\cite{AubryPRL1996,PiazzaPRL2007,Luccioli2011,Zambrini2013} since they usually emerge on the most connected residues~\cite{PiazzaPRL2007}, which are deeply buried inside the protein. On the other hand, surface damping can favor energy transfer phenomena as discussed in~\cite{PiazzaPRL2007,Luccioli2011}. This is indeed what we find both in Kinesin and Hemoglobin. Fig.~\ref{Fig_Friction} shows the results obtained for kinesin: we plot the energy of the first three NMs when an initial kick of energy $E_0=30$ Kcal/mol is given to the residue MET 96 (same initialization protocol as in Fig.~\ref{Fig_transfer}). The viscous friction coefficient is set to $\gamma = 0.2$ ps$^{-1}$. 
At the beginning NM3 (MET 96) loses energy, which flows across the whole protein structure and dissipates on the surface. At time $t\approx 50$ ps an energy transfer occurs with NM1 and NM2 (CYS 296 and THR 298) that harvest part of the energy from the system. Then also these NMs (residues) lose their energy but more slowly than NM3 (MET 96) and consequently also the total energy dissipation rate decreases. 
In fact, by fitting the total energy to an exponential function, before and after the energy transfer has taken place, one finds that the energy decay rate is smaller when most of the energy has been transferred from NM3 into NM1, see inset of fig.~\ref{Fig_Friction}. Interestingly, and similarly to what found for myoglobin in Ref.~\cite{PiazzaPRL2005}, the energy decay in the inset of Fig.~\ref{Fig_Friction} can be fitted to a single stretched exponential curve, that gives a good agreement over the whole time range considered here (fit not shown).
Eventually, at $t = 400$ ps NM 1 carries about $70$\% of the total energy.
This is another signature that energy transfer phenomena make the protein device thermodynamically efficient, by decelerating the dissipation. In fact, the transfer mechanism is such that the energy is harvested by modes which are ``buried'' inside the device structure (NM1 in this case), as we argue below.

In order to characterize the degree of contact with the solvent of the residues participating in the oscillation of a given NM, we introduce the quantity $\Phi_k$ for each NM $k$, where $\Phi_k=0$ signals that no residue participating in the normal mode motion interacts with the implicit solvent, and, conversely, $\Phi_k=1$ signals that all the residues participating in the normal mode motion are on the surface thus interacting with the solvent (the eigenvectors $\pmb{\xi}^k$ are normalized to one).
\begin{equation}
\Phi_k =  \left[ \sum_i \left( \sum_{\alpha} (\xi^k_{i,\alpha})^2 \right) \cdot \chi_i \right],
\label{eq:phi}
\end{equation}  
with $i=1,2,\ldots,N$, and  $\alpha=x,y,z$, and
where  $\chi_i = 1$ (0) if the residue $i$ is (is not) on the protein surface, i.e.,  if it can (cannot)  be touched by a geometrical sphere with radius $1.4$ \AA, rolling on the protein surface.

In the case of kinesin, sorting by increasing interaction with the solvent, we find
$\Phi_7=0.023$,
$\Phi_1=0.023$,
$\Phi_4=0.032$,
$\Phi_2=0.035$,
$\Phi_6=0.037$,
$\Phi_5=0.038$,
$\Phi_{10}=0.044$,
$\Phi_{26}=0.049$,
$\Phi_3=0.050$,
thus by transferring the energy from the NM3 into the NM1, the device optimizes the {\it storing} of such energy, by minimizing the dissipative interaction with the solvent.

\begin{figure}[htbp]
\center
\includegraphics[width=8cm]{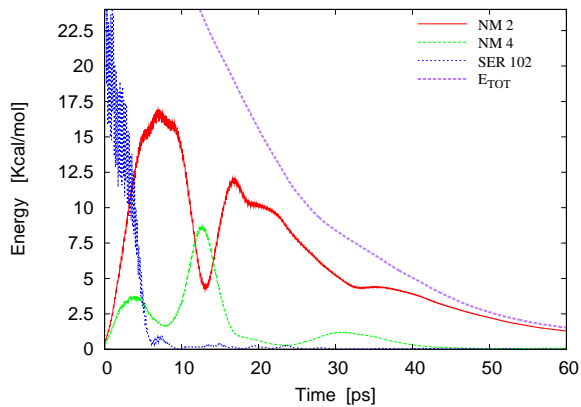}
\caption{(color online) Formation of DB in Hemoglobin after an initial kinetic kick of energy $E_0 = 50$ Kcal/mol at site SER 102 in chain A in presence surface friction ($\gamma = 0.2$ ps$^{-1}$). Energy of NM2, NM4 and total energy vs. time.}
\label{Fig_Friction_Hemo}
\end{figure}

A similar picture emerges for Hemoglobin in implicit solvent.
Fig.~\ref{Fig_Friction_Hemo} shows the energy of NM2, NM4 and the total energy when an initial kick with $E_0=50$ Kcal/mol is given to the binding site SER 102 in chain A (same initialization protocol as in Fig.~3 in Main Text). The viscous friction coefficient is set to $\gamma = 0.2$ ps$^{-1}$. From the very beginning, the energy flows from the NMs involved in the vibrations of SER 102 (NM48, NM29, NM43, and others ) into NM2 and NM4 and at $t=60$ ps NM2 carries almost the entire energy of the system. With $\Phi_2=0.282$ and $\Phi_4=0.425$,  NM2 and NM4 are respectively the first and the $19$-th most ``buried'' NMs. Interestingly, in both the simulated trajectory discussed here and in the corresponding trajectory without friction discussed in the Main Text, NM2 and NM4 are the most energetic NMs after energy transfer has occurred, but the transfer is more efficient in the presence of interaction with the solvent: indeed the transfer time is reduced (compare with fig.~(\ref{Fig_transfer_hemo})), and the energy is entirely transferred to NM2, which is the mode the less interacting with the explicit solvent.


In conclusion, our results show that non--linear phenomena are efficient mechanisms for energy transmission and storage in molecular devices, in three ways. There exists a range of activation energy, such that in this energy range, {\it i)} the transfer time is a minimum,  {\it ii)} the fraction of transferred energy is a maximum, and  {\it iii)} after the transfer, the energy is localized in modes which exhibit a minimal interaction with the solvent, thus minimizing the dissipation.

\medskip
\begin{acknowledgments}
We would like to thank F. Piazza for useful discussions. AI gratefully acknowledges financial support by the Danish Natural Science Research Council.
\end{acknowledgments}

\end{document}